\documentclass[twocolumn,showpacs,preprintnumbers,amsmath,amssymb]{revtex4}

\usepackage{graphicx}
\usepackage{dcolumn}
\usepackage{bm}

\begin{document}

\title{Nonlinear kinetic inductance sensor}

\author{D.~Yu.~Vodolazov}
\affiliation{Institute for Physics of Microstructures, Russian
Academy of Sciences, Nizhny Novgorod, GSP-105, Russia}

\email{vodolazov@impras.ru}

\begin{abstract}
The concept of nonlinear kinetic inductance sensor (NKIS) of
electromagnetic radiation is proposed. The idea is based on
divergency of kinetic inductance $L_k \sim dq/dI$ ($\hbar q$ is a
momentum of superconducting electrons, $I$ is a supercurrent) of
hybrid superconductor/normal metal (SN) bridge at current
$I^*<I_{dep}$ ($I_{dep}$ is a depairing current of the hybrid) and
temperature $T^*$ much smaller than critical temperature $T_c$. It
makes possible to have large change of phase difference $\delta
\phi$ along SN bridge in current biased regime at $I\simeq I^*$
even for small electron temperature increase. Appearance of
$\delta \phi$ is accompanied by the change of the current and
magnetic flux through the coupled superconducting ring which could
be measured with help of superconducting quantum interference
device (SQUID). In some respect proposed sensor may be considered
as a superconducting counterpart of transition edge sensor (TES)
those work is based on large derivative $dR/dT$ ($R$ is a
resistance) near $T_c$. Because at $I \simeq I^*$ SN bridge is in
gapless regime there is no low boundary for frequency of detected
e.m. radiation. Our calculations show that such a sensor can
operate in single photon regime and detect single photons with
frequency $\nu \gtrsim$ 10 GHz. We argue that the nontrivial
dependence $I(q)$ of SN bridge could be also used in detectors of
continuous e.m. radiation, current and magnetic field sensors.
\end{abstract}

\maketitle

\section{Introduction}

Superconductors are widely used in our days as a basic element of
detector/sensor of electromagnetic radiation. Absorbed photons
create quasiparticles and/or increase their energy in
superconductors and it changes its superconducting/transport
properties. For example in kinetic inductance detector (KID)
\cite{Zmuidzinas_2012} and transition edge sensor (TES)
\cite{Irwin_2005} kinetic inductance and resistance increases,
correspondingly, work of superconductor-insulator-superconductor
(SIS) mixer is based on change of its highly nonlinear
current-voltage characteristic \cite{Gundlach_2000}, while in
superconducting nano- and microstrip single photon detectors
(SSPD) current carrying superconducting strip switches to the
resisitive/normal state after appearance of hot spot (region with
heated electrons) in the place where photon is absorbed
\cite{Natarajan_2012}.

In case of TES its high sensitivity is based on narrow resistive
transition of superconductor $R(T)$ near critical temperature
$T_c$ and usage of superconducting quantum interference device
(SQUID) to measure response of the detector - the larger
derivative $dR/dT$ the larger resistance and current change via
voltage-biased superconductor. One of the advantages of TES is
that it operates in resistive state near $T_c$ and, hence,
superconducting gap practically vanishes, which means that there
is no low boundary typical for KID or SSPD where energy of the
photon $h\nu$ should exceed twice the superconducting energy gap
$\epsilon_g$. Besides it can work in single photon regime (for
infra-red and photons with larger energy) and resolve energy of
the absorbed photon contrary to SSPD.

\begin{figure}[hbtp]
\includegraphics[width=0.5\textwidth]{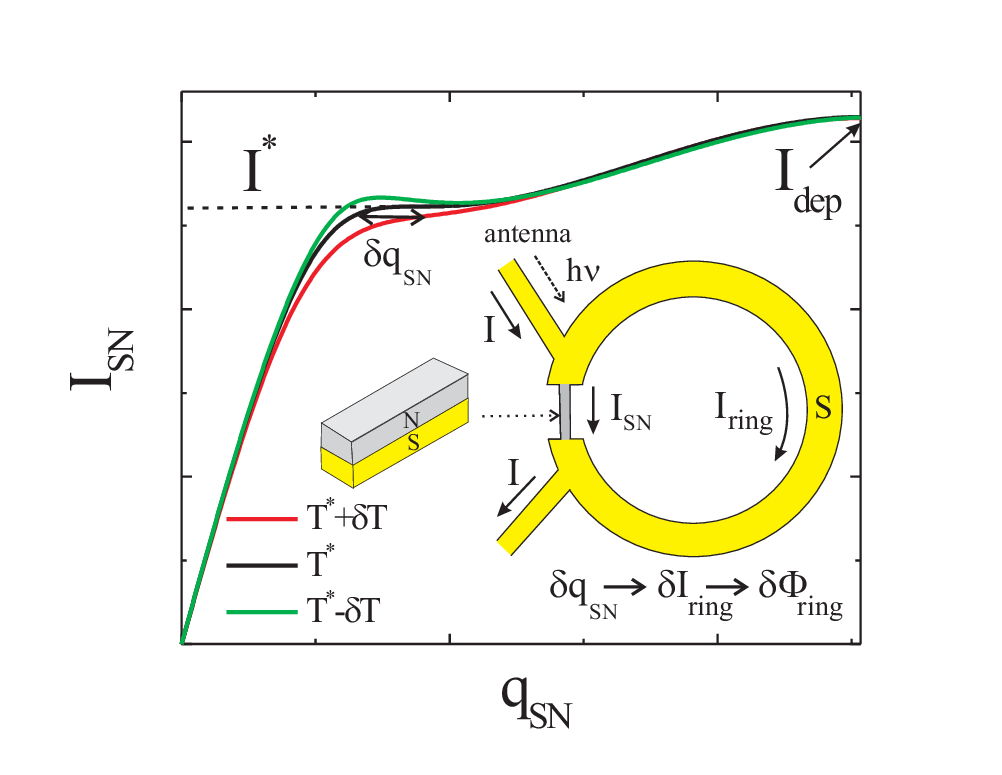}
\caption{Sketch of nonlinear kinetic inductance sensor (NKIS).
NKIS consists of SN bridge coupled with superconducting ring. With
decreasing of temperature on dependence of supercurrent on
momentum $I_{SN}(q_{SN})$ of SN bridge 'plateau' appears at
$T=T^*$ and $I_{SN}=I^*$ (marked by dashed line) which means a
divergency of kinetic inductance $L_k \sim dq_{SN}/dI_{SN}$. At
$I_{SN}\simeq I^*$ momentum changes by $\sim \delta q$ with
increase of electron temperature of SN bridge from $T=T^*$ up to
$T^*+\delta T$. It produces the change of the current in
superconducting ring and magnetic flux through it. The last could
be measured by SQUID. It is supposed that SN bridge and ring are
parts of antenna and photon with energy $h\nu$ less than twice
superconducting gap in superconducting electrodes can be absorbed
only by SN bridge where superconducting gap is equal to zero at
$I_{SN} \gtrsim I^*$.}
\end{figure}

Here we offer another type of energy resolved sensor which as a
TES is based on large variations of transport properties due to
small temperature increase of the whole superconducting detecting
element. Contrary to TES it operates in {\it superconducting}
state at $T \ll T_c$ but despite that it has zero superconducting
gap. The proposed nonlinear kinetic inductance sensor (NKIS)
consists of current biased hybrid superconductor-normal metal (SN)
bridge coupled to superconducting ring (see Fig. 1). SN bridge has
unique dependence of supercurrent $I_{SN}$ (throughout the paper
we use absolute value of the current) on momentum $\hbar q_{SN}$.
At $T=T^* \ll T_c$ there is a 'plateau' on dependence
$I_{SN}(q_{SN})$ at $I_{SN}=I^*$ where kinetic inductance
$L_k=l_{SN}\hbar (dq_{SN}/dI_{SN})/2|e|$ diverges ($l_{SN}$ is a
length of SN bridge). At $T>T*$ 'plateau' transforms to part of
$I_{SN}(q_{SN})$ with finite slope and finite $L_k$. Therefore one
may have large change of $q_{SN}$ and phase difference $\delta
\phi$ by biasing of SN bridge at $I_{SN}\simeq I*$ and bath
temperature $T=T^*$ even in case of small increase of electron
temperature. Increase of $q_{SN}$ leads to increase of the current
in superconducting ring and magnetic flux through it. The last
change could be measured by SQUID. We assume that SN bridge and
ring are parts of the antenna which receives the photon and photon
can be absorbed only in SN bridge which is in gapless regime at $I
\gtrsim I^*$ and in the rest of the system $h\nu<2\epsilon_g$.

In comparison with TES where the large derivative $dR/dT$ provides
large change of the current, in NKIS the large derivative
$dq_{SN}/dI_{SN}$ provides the large change of $I_{ring}$.  In
contrast to TES proposed NKIS operates in superconducting,
dissipation less state, and $dq_{SN}/dI_{SN} \to \infty$ at
$I_{SN}=I^*$ and $T=T^*$ which potentially may lead to larger
sensitivity than TES has. Like a TES it also does not have low
boundary for photon detection because at current $I_{SN} \gtrsim
I^*$ the superconducting gap is equal to zero in SN bridge. Below
we show that NKIS can potentially detect even single photons with
$\nu >$ 10-30 GHz when it is biased at temperature $15-150 mK$.

\section{Nonlinear kinetic inductance sensor}

In this section we present results of our calculations which
demonstrate detection ability of NKIS in single photon regime. In
Fig. 2(a) we show dependence $I_{SN}(q_{SN})$ at different
temperatures for SN hybrid bridge having following parameters:
thicknesses of S and N layers $d_S=3\xi_c$, $d_N=1.5 \xi_c$
($\xi_c=(\hbar D_S/k_BT_{c0})^{1/2}$), ratio of diffusion
coefficients $D_N/D_S=20$, $\hbar q=\hbar (\nabla \varphi+2\pi
A/\Phi_0)$ is the momentum of Cooper pairs ($\varphi$ is the phase
of superconducting order parameter, A is a vector potential,
$\Phi_0$ is a magnetic flux quantum), $T_{c0}$ is the critical
temperature and $I_{dep,S}$ is a depairing current at $T=0$ of
single S layer. To find it we use one-dimensional Usadel model
(equations and method of calculation are present in
\cite{Marychev_2021}) where we take into account dependence of
superconducting properties only across thickness of SN bridge
$d_S+d_N$ and assume that along its length ($l_{SN} \gg w_{SN}$)
and across the width ($w_{SN}<\Lambda$, $\Lambda$ is a Pearl
penetration depth) current density is distributed uniformly.

At temperature near $T^*$ on dependence $I_{SN}(q_{SN})$ there is
a 'plateau' at current $I=I^*$. It occurs due to temperature
driven transition from $I_{SN}(q_{SN})$ having one maximum at high
T to dependence having two maxima at low T. Maximum at low
$q_{SN}$ appears due to increased contribution of N layer to total
supercurrent (proximity induced superconductiviy in N layer
becomes 'stronger'), which also leads to increase of the slope
$I_{SN}(q_{SN})$ at small $q_{SN}$ and smaller kinetic inductance.
Because N layer has larger diffusion coefficient the depairing
effect of supervelocity (in Usadel equation it is described by
term with $\hbar D_{S,N}q_{SN}^2$) is larger in N layer than in S
layer ($q_{SN}=const$ across the thickness of SN bridge). It leads
to suppression of superconductivity in N layer at smaller $q_{SN}$
than in S layer and need to increase momentum to keep the same
supercurrent which results in appearance of the 'plateau'.

\begin{figure}[hbtp]
\includegraphics[width=0.45\textwidth]{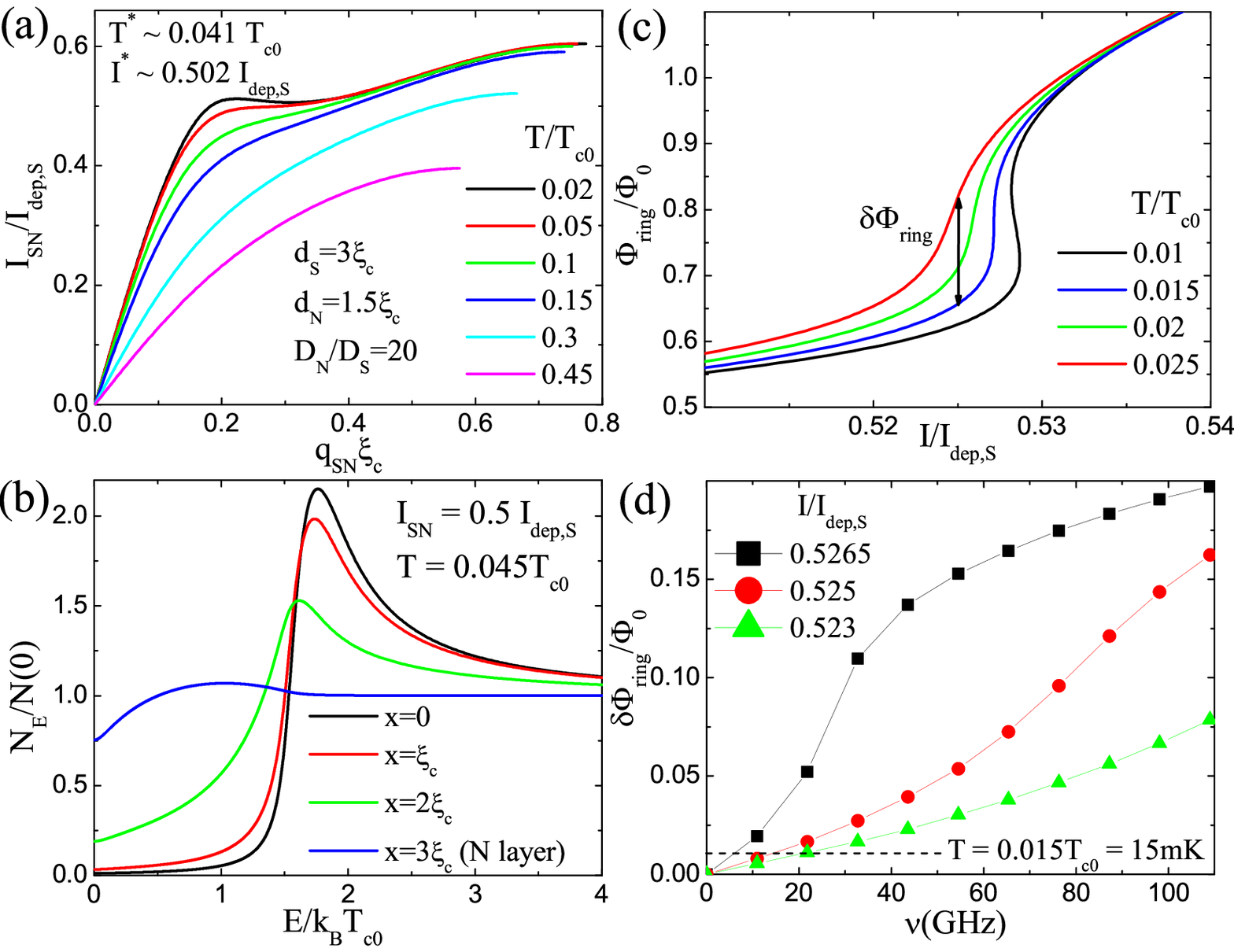}
\caption{(a) Dependence of supercurrent on momentum (stable branch
is shown with $dI_{SN}/dq_{SN}>0$) at different temperatures and
(b) density of states across SN bridge at $I_{SN} \sim I^*$ and $T
\simeq T^*$. (c) Dependence of the magnetic flux via
superconducting ring on current $I=I_{SN}+I_{ring}$ at different
temperatures. (d) Dependence of magnetic flux variation (see panel
(c)) on energy of the photon at different bias currents.}
\end{figure}

Fig. 2(b) demonstrates that at $I_{SN} \sim I^*$ the
superconductivity in SN bridge is gapless. We discuss gapless
regime in details below and here only notice that although gap is
zero but density of states in large part of S layer is much
smaller than in the normal state at $E \lesssim k_BT_{c0}$. We
estimate the effective 'normal' thickness of the SN bridge where
density of states is close to normal state value $N(0)$ as
$d_{eff}=d_N+\xi_c$.

When SN bridge is connected with a superconducting ring the total
current $I$ consists of $I_{SN}$ and $I_{ring}$ (see Fig. 1). We
assume that the ring is made of the same material as S layer of SN
bridge and it has the same thickness $d_S$ but its width is much
larger than the width of SN bridge: $w_{ring}\gg w_{SN}$. Because
of small thicknesses and widths of SN bridge and ring we may
neglect screening effects and assume that $\hbar q= \nabla \varphi
\gg A$. From condition that $\oint \nabla \varphi dl=0$ over the
ring perimeter and SN bridge (we assume there is no fluxoid inside
the ring) we find: $q_{ring}=q_{SN}l_{SN}/l_{ring}$. With this
relation and definition $\alpha=l_{SN}w_{ring}/l_{ring}w_{SN}$ we
may write expression for the total current:
\begin{equation}
I=I_{SN}(q_{SN})+1.55\alpha q_{SN}\xi_cI_{dep,S},
\end{equation}
where we use linear dependence $I_{ring} \sim q_{ring}\xi_c$
(because $q_{ring}\xi_c \ll 1$) and weak temperature dependence of
depairing current of S layer at $T \ll T_{c0}$.

With known dependence $I_{SN}(q_{SN})$ and expression for magnetic
flux through the ring $\Phi=L_GI_{ring}$ ($L_G=\mu_0R(ln(8R/r)-2)$
is geometric inductance of the ring, R is its radius, $r\sim
w_{ring}$ and $\mu_0$ is a magnetic constant) we find dependence
$\Phi(I,T)$. We show it in Fig. 2(c) for different temperatures
and following parameters: $w_{SN}=100 nm$, $l_{SN}= 1 \mu m$,
$w_{ring}=12 \mu m$, $l_{ring}=2.4 mm$, $D_S=0.5 cm^2/s$,
$T_{c0}=1 K$ ($\xi_c$=19.5 nm, $I_{dep,S}$=30,6 $\mu$A,
$\alpha=1/20$). Material parameters of superconductor are typical
for granular Al \cite{Levy-Bertrand_2019} and close to parameters
of other high resistive low temperature superconductors like TiN
\cite{Leduc_2010}, PtSi \cite{Szypryt_2016} or Hf
\cite{Coiffard_2020}. We also choose Cu as a material for N layer.

In the following we consider ability of NKIS detect single photons
(analysis can be also made when there is continuous flux of e.m.
radiation, as it was done in \cite{Giazotto_2008}). Using above
results we find dependence $\delta \Phi$ on change of the
temperature $\delta T$ of SN bridge after absorption of the single
photon with energy $h\nu$ - see Fig. 2(d). We relate $\delta T$
with $h\nu$ with help of energy conservation law
\begin{equation}
\delta T=\frac{h\nu}{C_eV_{eff}},
\end{equation}
where $C_e=2\pi^2k_B^2N(0)T_0/3$ is a heat capacity of electrons,
$T_0$ is bath temperature and $V_{eff}=l_{SN}w_{SN}d_{eff}$ is a
volume of SN bridge where density of electron states is close to
its normal state value. We also neglect heating of phonons due to
their much smaller heat capacity and use in calculations $N(0)= 13
eV/nm^3$ which corresponds to Cu.

The important characteristic of photon sensor is the signal to
noise ratio $\mathcal{S}/\mathcal{N}$. Noise may be estimated as
$\mathcal{N}=\delta \Phi_n \sqrt{\omega}$ where $\delta \Phi_n$ is
the flux sensitivity of the dc SQUID and $\sqrt{\omega}$ is its
bandwidth \cite{Giazotto_2008}. With $\delta \Phi_n =
10^{-6}/\sqrt{Hz}$ (which is ten times larger than the value
reported in \cite{Kiviranta_2004}) and $\omega=1 MHz$ we find
$\mathcal{N}=10^{-3}\Phi_0$. Therefore with a magnitude of signal
$\mathcal{S}=10^{-2}\Phi_0$ the ratio
$\mathcal{S}/\mathcal{N}=10$.

In Fig. 2(d) level $\mathcal{S}=10^{-2}\Phi_0$ is marked by dashed
line and our results demonstrate that proposed sensor could be
able to detect single photons with $\nu \gtrsim 10 GHz$ when it is
biased at $T = 15 mK$ and current $I \sim I^* \sim 0.526 I_{dep,S}
\sim 15 \mu A$.

The same sensor could be used to detect single THz and subTHz
photons. Because of their much larger energy we do not need mK
temperatures and $T_{c0}= 1 K$. Note that $V_{eff}\sim \xi_c \sim
1/\sqrt{T_{c0}}$ and $I_{dep,S} \sim T_{c0}$ for superconductor
those thickness scales with $\xi_c$. Therefore our results could
be easily rescaled - one needs to multiply $\delta\Phi/\Phi_0$ in
Fig. 2(d) by $T_{c0}(K)/1K$ and $\nu$ by $(T_{c0}(K)/1K)^{3/2}$.
For example for superconductor with $T_{c0}=10K$ (NbN, NbTiN)
absorption of photon with $\nu = 300 GHz$ at $T_0=150 mK$ leads to
change of magnetic flux by $\sim 0.2\Phi_0$ while 30 GHz photon
changes magnetic flux by $\delta \Phi/\Phi_0 \sim 0.02$.
\begin{figure}[hbtp]
\includegraphics[width=0.45\textwidth]{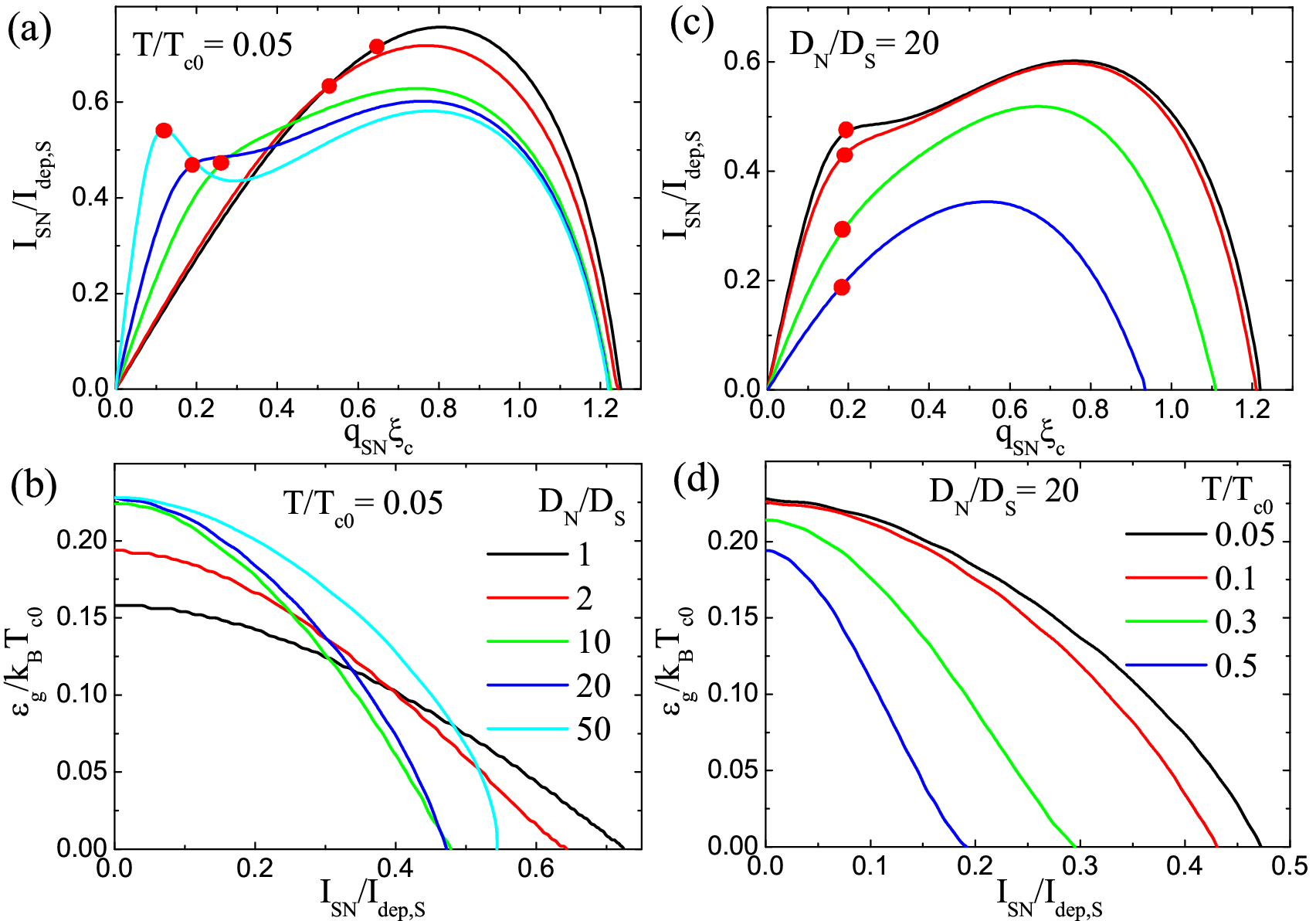}
\caption{(a) Dependence of superconducting current on momentum and
(b) current dependent superconducting gap in SN bilayer at fixed
temperature $T=0.05T_{c0}$ and different $D_N/D_S$. Red circles
indicate the value of the current $I_q$ and momentum $g_q$ where
gap vanishes. One can see that with increasing $D_N$ it occurs at
smaller $q_{SN}$. (c) Dependence of superconducting current on
momentum and (d) current dependent superconducting gap in SN
bridge at fixed $D_N/D_S=20$ and different temperatures.
Thicknesses $d_N$ and $d_S$ are the same as in Fig. 2.}
\end{figure}

After photon absorption electrons are heated on $\delta T$ and
than cool down due to electron-phonon interaction on time scale
$\sim \tau_{ep}$. In proposed system there is no cooling of
electrons by diffusion due to large energy gap on the ends of the
SN bridge where it is attached to superconductor with large gap
$\epsilon_g= 1.76 k_BT_{c0}$. At low temperature $\tau_{ep}\sim
1/T^3$ and for example for Ag it could be as large as
$10^{-1}-10^{-4} s$ in temperature range 10-100 mK
\cite{Giazotto_2008}. We expect that similar values are valid for
Cu too.

\section{Current controlled energy gap}

In this section we discuss gapless regime in SN bridge. In
ordinary 'dirty' superconductor energy gap is finite even at
$I=I_{dep}$ where it is about of $0.57 k_BT_{c0}\sim \Delta_0/3$
\cite{Maki_1963} ($\Delta_0\sim 1.76 k_BT_{c0}$ is the energy gap
at zero current). However in SN bridge gap goes to zero at
$I=I_g<I_{dep}$ which is marked by red circles in Fig. 3. Gapless
state in SN bridge is possible due to large difference between
diffusion coefficients in N and S layers. As we discuss earlier
pair breaking effect due to finite momentum in 'dirty'
superconductor is described by term $\sim \hbar Dq^2$ in Usadel
equation and when $\hbar Dq^2/2 \sim \Delta_0$ gap vanishes in
ordinary superconductor \cite{Maki_1963}. Roughly the same
criteria is valid in SN bilayer which one can see from Fig. 3(a)
where with increase of $D_N$ gap closes at smaller value of
momentum $q_{SN}$.

In ordinary superconductor condition $\hbar Dq^2/2 \sim \Delta_0$
is fulfilled on unstable branch of dependence $I(q)$ and
superconducting order parameter $\Delta$ rapidly decreases with
increase of $q$. In SN bilayer gapless state is stable due to
presence of S layer having much smaller diffusion coefficient and,
hence, much larger $q_g$.

\section{'Plateau' at different parameters of SN bridge}

We find that the 'plateau' exists in wide range of parameters of
SN bridge. With increase of $d_S$ (keeping $d_N$ and ratio
$D_N/D_S$ the same) temperature $T^*$, width of 'plateau' and
ratio $I^*/I_{dep}$ decrease (see panel (a) in Fig. 4). We also
find that there is minimal critical thickness $d_S$ (for example
it is about of $2.3 \xi_c$ for $d_N=1.5 \xi_c$ and $D_N/D_S=50$)
below which 'plateau' does not exist at any temperature. Panel (b)
illustrates that it is necessary to have large ratio $D_N/D_S
\gtrsim 20$ to have a 'plateau' (we find parameters when it exists
for $D_N/D_S=15$ but in this case $I^*\simeq I_{dep}$ which makes
impractical its usage). Another they to control $T^*$ is to change
thickness of N layer and finite interface transparency for
electron motion between S and N layers. For example increase of
$d_N$ or decrease of transparency shifts $T^*$ to lower
temperatures because in this case proximity induced
superconductivity in N layer becomes 'weaker' and one needs
smaller T to enhance it.

\begin{figure}[hbtp]
\includegraphics[width=0.45\textwidth]{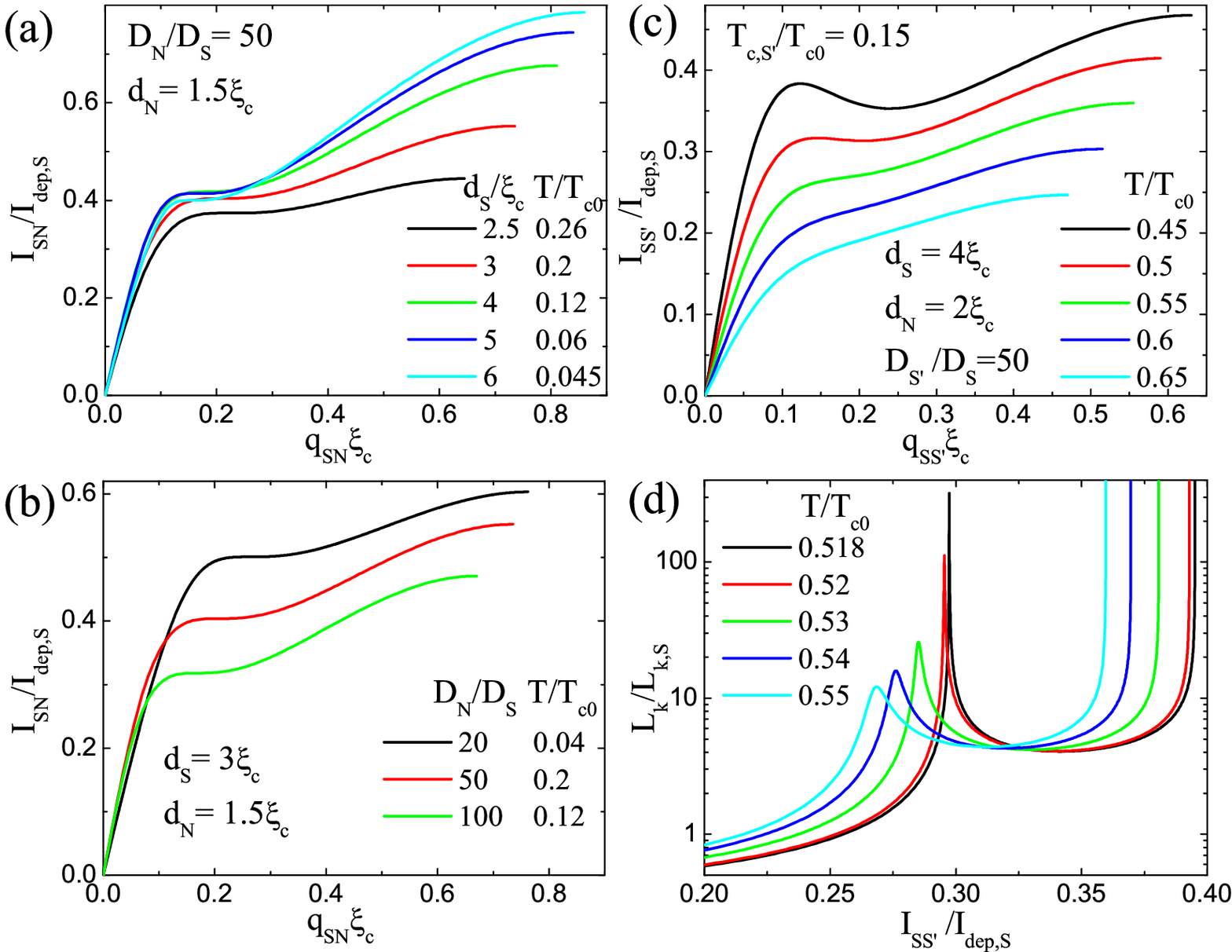}
\caption{(a) Dependence of superconducting current on momentum at
different $d_S$, fixed $d_N=1.5 \xi_c$, $D_N/D_S=50$ and
temperatures $T$ close to $T^*$. (b) Dependence $I_{SN}(q_{SN})$
at different $D_N/D_S$, fixed $d_N = 1.5 \xi_c$, $d_S = 3\xi_c$
and temperatures T close to $T^*$. (c) Dependence
$I_{SS'}(q_{SS'})$ for $SS'$ bridge at different temperatures and
$D_{S'}/D_S=50$, $d_N = 2 \xi_c$, $d_S = 4\xi_c$. (d) Current
dependent kinetic inductance of SS' bridge at different
temperatures close to $T^*$ ($L_{k,S}$ is a kinetic inductance of
single S layer at T=0).}
\end{figure}

The 'plateau' on $I(q)$ may also exist in bilayer composed of two
superconductors having different critical temperatures - see Fig.
4(c). Again gapless regime is realized at $I<I_{dep}$ and
dependence $I_{SS'}(q_{SS'})$ is sensitive to small variations of
temperature. The sensitivity to temperature is well visible in
Fig. 4(d) where we plot current dependent kinetic inductance. With
decrease of the temperature peak appears on dependence
$L_k(I_{SS'})$ which becomes sharp at $T=T^*$ and $I_{SS'}=I^*$.

\section{Discussion}

Peak on dependence $L_k(I)$ has been observed recently for
MoN(40nm)/Cu(40nm) strip ($d_S=d_N\simeq 6 \xi_c$, $T_{c0} \simeq
7.8K$) \cite{Levichev_2023}. With lowering of temperature width of
the peak decreases while amplitude increases in accordance with
prediction of Usadel model. In that experiment the critical
current was smaller than depairing current (most probably due to
edge defects) because divergency of $L_k$ has not been found at
$I=I_c$ \cite{Levichev_2023}. The thickness of studied sample was
optimized to observe finite momentum superconductivity at
relatively small in-plane magnetic field and narrow peak should
appear at $T \simeq 0.025 T_{c0} \simeq 200 mK$, as Usadel model
predicts, which is much lower than the minimal reached temperature
$T=2.7K$ in that experiment.

Single microwave photon detectors based on heating of small volume
superconductor have been proposed earlier. The similar design as
in Fig. 1 but with SNS Josephson junction (JJ) instead of SN
bridge was discussed in Ref. \cite{Giazotto_2008}. Authors of that
work offer to exploit strong temperature dependence of critical
current of long JJ and they predict that such a device would be
able to detect THz photons. In superconducting loop with two SNS
junctions heating of one JJ by absorbed photon produces voltage
pulse (when magnetic flux through the loop is close to $\Phi_0/2$)
and it is idea of another type of single THz photon detector
\cite{Solinas_2018}. In Ref. \cite{Paolucci_2023} more complicated
design was studied with two coupled superconducting loops (one
with magnetic flux $\sim \Phi_0/2$ and another with $\Phi=0$) and
two (short and long) SNS Josephson junctions. It is assumed that
microwave photon is absorbed in long JJ and it leads to the
measurable change of quasiparticle current via tunnel junction
connected with short JJ - for that device ability to detect
photons in range 10 GHz-10 THz was predicted. In
\cite{Astafiev_2022} the change of impedance of SN bridge placed
in the end of superconducting coplanar line serves as a signal
that microwave photon is absorbed there and this device does not
need external current drive.

In comparison with Refs.
\cite{Giazotto_2008,Solinas_2018,Paolucci_2023} our system has
more simple design and it is not based on usage of SNS junction.
As for the system from \cite{Astafiev_2022} it is possible to vary
parameters of SN bridge (width, length, thickness of N layer) to
match NKIS with 50 $\Omega$ microwave impedance of the coplanar
feed line when it is needed. Due to similarity with TES it is
possible to use experience and methods which has been developed
for that device. In comparison with TES the advantage of NKIS is
that it operates in superconducting state. But more important that
there is a current and temperature where $dI_{SN}/dq_{SN} \to 0$
and $dq_{SN}/dI_{SN} \sim L_k \to \infty$ at $I=I^*<I_{dep}$ which
is consequence of temperature driven 'phase transition' from
dependence $I(q)$ having one maximum to $I(q)$ having two maxima
and appearance of the extremum at small $q$ near which $dI/dq$
changes sign.

In reality peak in dependence $L_k(I)$ will be smeared due to
temperature/current fluctuations, local variations of
thickness/width and/or material parameters (local $T_c$,
resistivity, etc.) of the sample which are also responsible for
broadening of resistive transition in TES. However in contrast to
resistive transition where always $dR/dT>0$ and $R(T)$ is
reversible on dependence $I(q)$ at $T<T*$ there is a region with
$dq/dI<0$ near $I^*$ (see Fig. 1). It implies hysteretic behavior
of $L_k(I)$ near $I^*$. If we choose temperature slightly below
$T^*$ increase of electron temperature leads to jump of $L_k$ and
$\delta \phi$ which is easy to detect. But after cooling of
electrons SN bridge does not return to initial state which is not
convenient for following detection process. In this case
fluctuations help to come back to the initial state while change
of $\delta \phi$ remains large. In some respect situation is
similar to current-voltage characteristics (IVC) of
superconducting bridge. Usually at $T\sim T_c$ IVC is smooth and
reversible while at $T\ll T_c$ it is hysteretic with abrupt
voltage jump at $I=I_c$. Somewhere in between there is temperature
when IVC is reversible but with large $dV/dI$ at $I \simeq I_c$.
We expect something similar effect at $T \sim T^*$ but for
dependencies $q(I)$ and $dq/dI$ near $I^*$.

Although we mainly concentrate on ability to detect single
microwave photon the NKIS can also work in continuous regime like
TES (hot electron bolometer). In this case to find $\delta T$ one
has to solve balance equation which includes heating of SN bridge
due to absorbed e.m. radiation and its cooling due to
electron-phonon interaction. Note, that for this type of
application is not necessary to use SQUID readout and small size
SN bridge. As in KID one may use coplanar waveguide with
capacitively coupled superconducting circuit which contains
current biased SN bridge. Biased at $I=I^*$ and $T=T^*$ SN bridge
drastically changes its kinetic inductance when temperature
increases (see evolution of $L_k(I)$ with temperature in Fig.
4(d)). Due to gapless superconductivity quality factor of a
superconducting circuit should be small in comparison with
ordinary KID which means broadening of the resonance but it could
be compensated by extremely large change of $L_k$.

Divergency of $L_k$ at $I\to I^*$ could be also used in current
and magnetic field sensors. Indeed, even small deviation of the
current from $I^*$ leads to large change of $L_k$ at temperature
close to $T^*$ - see Fig. 4(d). This property could be used in
kinetic inductance parametric up-converter \cite{Kher_2016} to
amplify small current signal in TES instead of SQUID and kinetic
inductance magnetometer \cite{Luomahaara_2014} to measure
screening current produced by magnetic field.

\begin{acknowledgments}
The work is supported by the Ministry of Science and Higher
Education of the Russian Federation (contract no. 075-15-2022-316
with the Photonics Center of Excellence).
\end{acknowledgments}

\end{document}